\documentstyle[prl,epsf,eqsecnum,aps]{revtex}
\def\be{\begin{equation}}
\def\ee{\end{equation}}
\def\bea{\begin{eqnarray}}
\def\eea{\end{eqnarray}}
\begin{document}
\draft
\preprint{\parbox{1.5in}{ \leftline{WM-99-110}
                            \leftline{JLAB-THY-99- }
                         }
          }
\title{Confinement and the analytic structure of the one body propagator in
Scalar QED}
\author{\c{C}etin \c{S}avkl{\i}$^{1}$, Franz Gross$^{1,2}$, John Tjon$^{3}$  }
\address{
$^1$Department of Physics, College of William and Mary, Williamsburg,
Virginia 23187\\
$^2$Jefferson Lab,
12000 Jefferson Avenue, Newport News, VA 23606\\
$^3$Institute for Theoretical Physics, University of Utrecht,
Princetonplein 5,\\
P.O. Box 80.006, 3508 TA Utrecht, the Netherlands.}
\date{\today}
\maketitle

\begin{abstract}
We investigate the behavior of the one body propagator in SQED. The self
energy is calculated using three different methods: i) the simple
bubble summation, ii) the Dyson-Schwinger equation, and iii) the
Feynman-Schwinger represantation. The Feynman-Schwinger representation
allows an {\em exact} analytical result. It is shown that, while the exact 
result produces a real mass pole for all couplings, the bubble sum and the 
Dyson-Schwinger approach in rainbow approximation leads to complex mass poles 
beyond a certain critical coupling.  The model exhibits confinement, yet the 
exact solution still has one body propagators with {\it real\/} mass poles.
\end{abstract}


\narrowtext
\twocolumn

\section{Introduction}
\label{introduction}
The nature and implications of particle confinement remain one of the
mysteries of QCD.  It is
clear that confinement implies that quarks and antiquarks cannot be
separated from
each other at large distances (as demonstrated by lattice
calculations\cite{lattice1,lattice2}).  An
essential consequence of this is that a bound state cannot decay into its
constituent quarks
even if the decay is kinematically allowed.  Such a decay will certainly be
prevented if the
dressed quark propagators cannot have any real mass poles. This possibility
has been
investigated, and often implicitly assumed, within the context of
Dyson-Schwinger Equations \cite{STINGL,REF2,KREWALD,ROBERTS1,REF3,SAVKLI}.
However, this condition is not necessary; an alternative
point of view is that the confinement is not due to the lack of mass poles
but through the
exchange interaction between the constituents forming the bound
state\cite{GROSS1,GROSS2,SAVKLI2}.  In Ref.~\cite{SAVKLI2}, the authors
have shown that a
relativistic generalization of the nonrelativistic  linear interaction
leads to a bound state
vertex function that vanishes  when both particles are on shell. According
to this result, the
correct nonrelativistic limit favors the {\em vanishing of the vertex
function when both
particles are on-shell}, rather than the {\em lack of physical mass  poles}.

\begin{figure}
\begin{center}
\mbox{
    \epsfxsize=3.2in
\epsffile{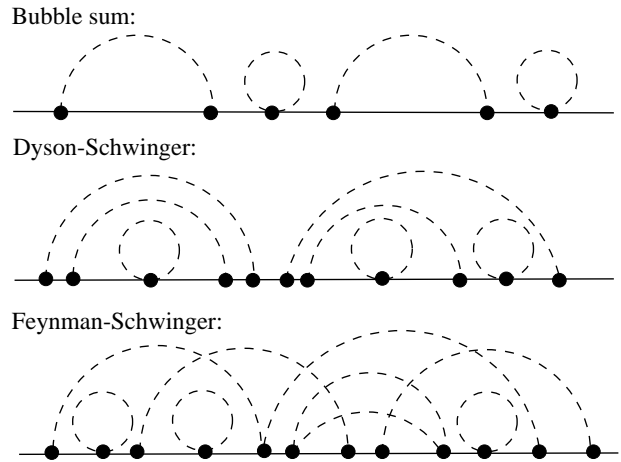}
}
\end{center}
\caption{ Various interactions included in each approach are shown.
The Feynman-Schwinger approach includes all diagrams. In 1-dimension the 
contribution of diagrams with loops of charged particles identically vanishes 
(Eq.~\ref{loops}). }
\label{allthree.sqed}
\end{figure}

Clearly the structure of the one-body propagator deserves a closer look and a
more rigorous understanding is needed to clarify what the Dyson-Schwinger
results mean. In this letter we study the one-body propagator in the
context of massive scalar QED in 0+1 dimension\cite{TJON2}.  The 
simplicity of this toy
model
field theory allows one to obtain an analytical solution for the dressed
mass by  using the
Feynman-Schwinger Representation
(FSR)~\cite{TJON2,SIMONOV1,TJON1,SAVKLI3,BRAMBILLA}. The FSR is
an approach based on Euclidean path integrals similar to lattice gauge
theory. In  this
approach the path integrals over quantum fields are integrated out at the
expense of introducing
path integrals over the trajectories of the particles.  The FSR
approach sums up all possible interactions including the ones with
``quark'' loops. Therefore
the FSR approach provides us the means to test and understand how much of
the physics is
included  in the Dyson-Schwinger equation with rainbow approximation. The
rainbow  approximation
corresponds to using bare interaction vertices and a bare  exchange field
propagator. At the
other  extreme, one may consider the dressed mass as obtained by a simple
bubble  summation.
This  method sums fewer diagrams than the other two. In
Fig.~\ref{allthree.sqed} the typical diagrams involved in all three
approaches are displayed.
In the next section we briefly discuss how the
dressed mass is obtained in each one of the three methods mentioned above,
and how the results compare with each other.
%
%
\section{Scalar QED}

Massive scalar QED in 0+1 dimension is a simple interaction that enables one
to obtain a fully analytical result for the dressed and bound state masses
within the FSR approach.
In this section we compare the self energy result obtained by three different
approaches; namely the simple bubble sum, the Dyson-Schwinger equation, and
the Feynman-Schwinger representation.
The Minkowski metric expression for the scalar QED Lagrangian in Feynman
gauge is given by
\begin{eqnarray}
{\cal
L}_{SQED}&=&-m^2\chi^2-\frac{1}{4}F^2+\frac{1}{2}\mu^2A^2-\frac{1}{2}( 
\partial A
)^2\nonumber\\
&+&(\partial_\mu-ieA_\mu)\chi^*(\partial^\mu+ieA^\mu)\chi,
\end{eqnarray}
where $A$ represents the gauge field of mass $\mu$, and $\chi$ is
the charged field of mass $m$.  The field tensor $F$ is zero in 0+1
dimensions, and the
dynamics is described by the gauge fixing term $(\partial A)^2$.  The
presence of a mass term for
the exchange  field breaks the gauge invariance. Here the mass term was
introduced in order  to
avoid infrared singularities which are present in 0+1 dimension. For
dimensions larger than
$n=2$ the infrared singularity does not exist and  therefore the limit
$\mu\rightarrow 0$ can be
safely taken to restore the gauge  invariance.

\subsection{The bubble sum}
The bubble sum is the simplest subset of all diagrams contributing to the
self energy.
The Euclidean expression for self energy in 0+1 dimension is given
  by
\begin{eqnarray}
\Sigma(p)&=&-e^2\int_{-\infty}^{\infty}
\frac{dk}{2\pi}\frac{(2p-k)^2}{(k^2+\mu^2)[(p-k)^2+m^2]}\nonumber\\
&&+e^2\int_{\infty}^{\infty}\frac{dk}{2\pi}\frac{1}{k^2+\mu^2}
\label{bubblesum}
\end{eqnarray}

The dressed propagator corresponding to this self energy is
\begin{equation}
\Delta_d(p)=\frac{1}{p^2+m^2+\Sigma_E(p)}\, .
\end{equation}
The dependence of
$M$ on the coupling strength $e$ can be obtained from the solution of the
on-shell condition
\be
M=\sqrt{m^2+\Sigma_E(iM)},
\label{shellcondition}
\ee
which must be real if the dressed mass is to be stable.
Therefore, for massive SQED, the equation determining the dressed mass
takes the
following form
\begin{eqnarray}
M^2=m^2&+&\frac{e^2}{2}\biggl[\frac{(\mu-2M)^2}{\mu[m^2-(\mu-M)^2]}\nonumber\\
&&\,\,\,\,\,\,\,\,\,\,\,\,+\frac{(m-M)^2}{m[\mu^2-(m+M)^2]}+\frac{1}{\mu}\biggr]
.\label{sqed.pert.mass}
\end{eqnarray}

\subsection{The Dyson-Schwinger equation}
The Dyson-Schwinger Equation is usually solved in the rainbow approximation.
This is due to the fact that a completely self consistent determination
of the interaction vertex is impossible. The one body Dyson-Schwinger
equation in rainbow approximation is given by
\begin{eqnarray}
m^2(p)&=m^2&-e^2\int_{-\infty}^{\infty}
\frac{dk}{2\pi}\frac{(2p-k)^2}{(k^2+\mu^2)[(p-k)^2+m^2(k)]}\nonumber\\
&&+e^2\int_{\infty}^{\infty}\frac{dk}{2\pi}\frac{1}{k^2+\mu^2},
\end{eqnarray}
The structure of this equation is very similar to the earlier bubble sum
expression Eq.~(\ref{bubblesum}). The main difference is the momentum
dependence of the dressed mass. The coordinate space form of the dressed
propagator is
\be
\Delta_d(t)=\int_{-\infty}^\infty
\frac{dp}{2\pi}\left(\frac{e^{ipt}}{p^2+m^2(p)}\right)
\simeq N e^{-Mt}\,  .
\ee
Therefore the ground state mass pole of the one-body propagator can be
extracted using
\begin{equation}
M=-\lim_{T\rightarrow \infty}\frac{d}{dT} {\rm log}[\Delta_d(t)].
\end{equation}

\subsection{The Feynman-Schwinger representation}
In the FSR approach the field theoretical path integral expression for the
one-body propagator is transformed into a quantum mechanical path integral
over trajectories of the particles~\cite{SIMONOV1,SAVKLI3}. The FSR expression
for the one body propagator is given by
\begin{equation}
G(0,T)=\int ds\int ({\cal D}z)_{0T}\, {\rm exp}\biggl[iK[z,s]-V[z]\biggr],
\label{gsqed}
\end{equation}
where
\bea
&&K[z,s]=(m^2+i\epsilon)s-\frac{1}{4s}\int_0^1 d\tau
\,\dot{z}^2(\tau),\label{k1}\\
&&V[z]=\frac{e^2}{2}\int_0^1 d\tau\,\dot{z}(\tau)\int_0^1 d\tau'
\,\dot{z}(\tau')\,\Delta(z(\tau)-z(\tau')),
\label{v.sqed}\\
&&\Delta(z)= \int \frac{dp}{2\pi}\frac{e^{ip
x}}{p^2+\mu^2}=\frac{e^{-\mu|z|}}{2\mu},
\label{kernel.sqed}
\eea
where $\Delta(z)$ is the interaction kernel, and the boundary conditions are 
chosen to be $z(0)=0$, and $z(1)=T$. $K[z,s]$ represents the
mass term and the kinetic term, and $V[z,s]$ is the interaction term.
Due to the simplicity of working in 1 dimension, the integral of the self
interaction
Eq.~(\ref{v.sqed}) can be done analytically
\begin{eqnarray}
V[z]&=&\frac{e^2T}{2\mu^2}\biggl[1-\frac{1-e^{-\mu T}}{\mu T}\biggr].
\end{eqnarray}
In higher dimensions the result of this integral depends on the
trajectory of the particle. However in 1 dimension all trajectories
contribute equally, which is what makes the 1 dimension calculation
analytically do able. In addition, the contribution of loops, which has been 
omitted in Eq.~\ref{gsqed}, can be shown to identically vanish in 1-dimension.
The typical loop contribution in 1-dimension has the following form
\begin{eqnarray}
&&\int_0^1 d\tau\,\dot{z}(\tau) \oint d\tau'\,\dot{z}(\tau') \Delta(z(\tau)-z(\tau'))\nonumber\\
&&=\int_0^Tdz\,\biggl(\int_{z_i}^{z_f}dz'+\int_{z_f}^{z_i} dz'\biggr)\Delta(z-z')=0\label{loops}
\end{eqnarray}
Therefore in 1-dimension matter loops do not contribute and the FSR results
provided here are exact. Next, the path integral over $z$ can be evaluated 
after a discretization in proper time. Since the only path dependence in the 
propagator is in the
kinetic term, the path integral over $z$ involves gaussian integrals which
can be
performed easily by using the following discretization
\be
({\cal D})_{0T}\rightarrow(N/4\pi s)^{N/2}\Pi^{N-1}_{i=1}\int dz_i\, .
\label{discreet}
\ee
The $s$ integral can be evaluated by the saddle point method giving
\begin{equation}
G(0,T)=N {\rm
exp}\biggl[-mT-e^2\frac{T}{2\mu^2}+\frac{e^2}{2\mu^3}(1-e^{-\mu T})\biggr].
\end{equation}
This is an exact result for large times $T$. The
dressed mass can easily be obtained by taking the logarithmic derivative of
this expression. Therefore,
the one-body dressed mass for SQED in 0+1 dimension according to the FSR
formalism is given by
\begin{equation}
M=m+\frac{e^2}{2\mu^2}.
\label{sqed.fsr.mass}
\end{equation}
Having outlined the calculation of the dressed mass in three different
approaches, we next compare the results obtained by these methods.

\section{Discussions and Conclusions}

The kind of diagrams included in each method discussed above is displayed in
Fig.~\ref{allthree.sqed}. The main difference between the
Dyson-Schwinger and the Feynman-Schwinger diagrams is the crossed
diagrams. These diagrams involve photon lines that cross each other. The FSR
approach also includes all possible four-point interaction contributions
while the rainbow DSE only includes the tadpole type four-point interactions.
In principle all four-point interactions can also be incorporated into the
simple bubble sum and the rainbow DSE.

\begin{figure}
\begin{center}
\mbox{
    \epsfxsize=3.5in
\epsffile{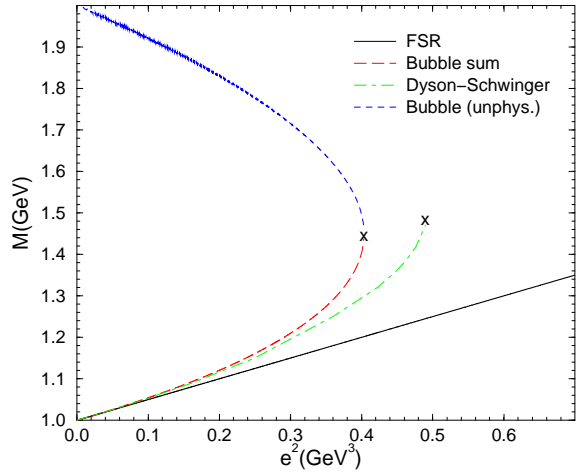}
}
\end{center}
\caption{The function $M(g^2)$ calculated by the FSR approach, the
Dyson-Schwinger equation,
and the bubble summation for values of $m=\mu=1$ GeV. According  to the
bubble sum there is a
critical point  at $e^2_{crit}=0.4$ (GeV)$^3$ beyond which the dressed
mass  becomes complex.
A similar result happens for the DSE. The FSR result is real for all
couplings.}
\label{mvsg2.sqed}
\end{figure}

In Fig.~\ref{mvsg2.sqed} we display all dressed mass results. The bubble
summation develops a complex mass pole beyond a critical coupling
$e^2_{crit}=0.4$ (GeV)$^3$.
At the critical point a  `collision' takes place with another real solution of
Eq.~(\ref{sqed.pert.mass}), leading to two complex conjugated solutions
with increasing $e^2$.
This  happens at $M=1.45$ GeV. It is interesting to note that the result
obtained from the
Dyson-Schwinger Equation displays a similar characteristic. At low
coupling strengths the
rainbow DS and the bubble results are very  close and they converge to the
exact result
given by the Feynman-Schwinger  approach. As the coupling strength is
increased the DS result
maintains a  closer distance to the bubble result rather than the FSR
result. Similar  to
the bubble result the DS result develops a complex mass pole at a
critical coupling of
$e^2_{crit}=0.49$ (GeV)$^3$.
%

There are two important observations to be made from these results. (i)
The dynamical generation of complex mass poles in the rainbow DS and
bubble approaches is not an indication of confinement (see also 
Ref.~\cite{AB78}).
These complex masses occur at large couplings, when it might appear 
that some sort of
confining phase
transition has taken place, but since the exact FSR answer shows no such
behavior we are forced
to conclude that these complex poles occur simply because the subset of the
possible interaction
diagrams included in these approaches is insufficient to qualitatively
reproduce the correct
result. (ii) The nature of the rainbow  DS result is closer to the
bubble sum than the exact FSR result.

Further insight follows from examination of the masses of two-body bound
states. The simplicity
of SQED in 0+1 dimension also allows one to get an analytical result for
the two-body bound state
mass. The total result is
\begin{equation}
M_{b}=\left(m+\frac{e^2}{2\mu^2}\right)+\left(m+\frac{e^2}{2\mu^2}\right)
-\frac{e^2}{\mu^2}=2m,
\label{sqed.fsr.bmass}
\end{equation}
where the first two terms are the dressed one-body contributions and the
last term is the contribution from the exchange interaction~\cite{TJON2}.
Hence there is {\it
only one\/} 2-body bound state, and the  {\em continuum spectrum does not 
exist}.
In light of the fact
that we are working in only 1 time  dimension the lack of a continuum is
not surprizing, since
the particles cannot move. In 0+1 time dimensions one would naturally
expect confinement simply
because there is no room for quarks to break free. However it is still
interesting to note that
confinement, which is unavoidable in 0+1 dimension, is a basic property
of the four-point function and it does not imply the lack of physical
mass poles in the one body propagator. This feature is similar to that
used in Refs.~\cite{GROSS2,SAVKLI2}.
Moreover, in QCD it would clearly also be more appropriate to discuss
confinement in the color-white sectors, e.g. for the $q{\bar q}$ system,
instead of for the single quark propagator.

The dynamical mass generation and binding contained in
Eq.~(\ref{sqed.fsr.bmass}) is
quantitatively similar to that  of the generation of massless Goldstone
bosons in QCD. In
particular it is  known that the pion mass $m_\pi$ is proportional to the
current quark mass
$m_u$,
\be
m_\pi^2=m_{u}\frac{<\bar{\Psi}\Psi>}{f_\pi},
\ee
where $<\bar{\Psi}\Psi>$ is the quark condensate and $f_\pi$ is the pion
decay constant.
This is similar to the result found in Eq.~(\ref{sqed.fsr.bmass}).
In SQED the positive shifts of one-body masses are exactly compensated by
the negative binding energy created by the exchange interaction. Therefore
the total bound state mass is exactly equal to the sum of {\em bare} masses,
and the bound state mass vanishes as the current particle mass vanishes.
However in scalar QED particles do not carry spin. Therefore the similarity
to the dynamical chiral symmetry breaking of QCD is only accidental.

Finally, the FSR formalism allows us to make the following observation
about the
significance of the vertex dressings of the interaction: If one starts with
dressed masses given
in  Eq.~(\ref{sqed.fsr.mass}) and uses only the exchange interaction to
calculate
the bound state masses, the resultant bound state mass would have been the
same. This means that the vertex contributions do not change the bound state
energy. This type of prediction underlines the potential usefulness of the FSR
calculations. In principle, besides being a rigorous and powerful tool for
calculation of the nonperturbative propagators, the FSR approach can also
provide much needed information about the role of various vertices and
propagators. This information would be useful as input in other
nonperturbative approaches such as the Dyson-Schwinger equations.

This paper focused on a simple toy model, namely the SQED in 0+1 dimension.
Through this simple model we have been able to compare various nonperturbative
methods. It would be interesting to see whether insights provided by this 
simple model could be extended to higher dimensions. We have in particular 
shown, that the exact solution is confining and yet the 1-body propagator
has real mass poles. Whether this scenario is realized in 3+1 dimension is an 
interesting and important question.
 
\section{acknowledgement}
This work was supported in part by the US Department
of Energy under grant No.~DE-FG02-97ER41032.

\end{document}